\documentclass[twocolumn,english,reprint,superscriptaddress,prl]{revtex4-1}

\usepackage[T1]{fontenc}
\usepackage[latin9]{inputenc}
\usepackage{color}
\usepackage{bm}
\usepackage{amstext}
\usepackage{amssymb}
\usepackage{graphicx}
\usepackage{booktabs}
\usepackage{tabularx}
\usepackage{amsmath}
\makeatletter

\usepackage{babel}

\usepackage{hyperref}
\hypersetup{
 linktocpage = true,
     colorlinks,
    citecolor=blue,
    filecolor=black,
    linkcolor=blue,
    urlcolor=blue
}

\usepackage{wasysym}

\makeatother

\begin{document}
\author{Giacomo Torlai}
\affiliation{Center for Computational Quantum Physics, Flatiron Institute, New York, New York, 10010, USA}
\author{Roger G. Melko}  
\affiliation{Department of Physics and Astronomy, University of Waterloo, Ontario, N2L 3G1, Canada}
\affiliation{Perimeter Institute for Theoretical Physics, Waterloo, Ontario N2L 2Y5, Canada}

\title{Machine learning quantum states in the NISQ era}

\begin{abstract}
We review the development of generative modeling techniques in machine learning for the purpose of reconstructing real, noisy, many-qubit quantum states.
Motivated by its interpretability and utility, we discuss in detail the theory of the restricted Boltzmann machine.
We demonstrate its practical use for state reconstruction, starting from a classical thermal distribution of Ising spins,
then moving systematically through increasingly complex pure and mixed quantum states.  Intended for use on  
experimental noisy intermediate-scale quantum (NISQ) devices, we review recent efforts in reconstruction of a cold atom wavefunction. 
Finally, we discuss the outlook for future experimental state reconstruction using machine learning, in the NISQ era and beyond.
\end{abstract}

\maketitle

\selectlanguage{english}

\section{Introduction}

We are entering the age where quantum computers with tens -- or soon hundreds -- of qubits are becoming available.  
These noisy intermediate-scale quantum (NISQ) devices \cite{NISQ} are being constructed out of cold atoms~\cite{Bernien2017}, superconducting quantum circuits~\cite{ibm_vqe}, trapped ions~\cite{monroe53}, 
and other quantum systems for which we have achieved an exquisite degree of control.
NISQ devices will soon play an important role, since they are poised to surpass the ability of the world's most powerful computers to perform exact simulations of them,
ushering in the era of so-called quantum supremacy \cite{QSupreme}.   

Some of the first tasks for these NISQ devices will be as simulators (or emulators) of other highly-entangled quantum many-body systems.
The goal is to supplant our current conventional computer simulation technology, such as exact diagonalization, quantum Monte Carlo, or tensor network methods.
Efforts to produce quantum simulators of some of the most important physical systems, such as the fermionic Hubbard model \cite{Hubbard}, are progressing in earnest.
However, with the advent of increasingly larger NISQ devices comes a paradox: how will we simulate the simulators?  That is, how will we validate  
an intermediate-scale quantum device, confirming that it is producing the behavior it was designed for?  
Along with quantum supremacy comes the necessary breakdown of conventional tomography -- the gold standard for quantum state reconstruction.  We are left searching for imperfect alternatives.

The answer may lie in new data-driven approaches inspired by rapid advances in machine learning. 
A strategy for unsupervised learning, called {\it generative modeling}, has demonstrated the ability to 
integrate well with the data produced by NISQ devices.  In industry applications, the goal of generative modeling is to reconstruct an unknown probability 
distribution $P(\bm{x})$ from a set of data $\bm{x}$ drawn from it.  In the most powerful versions of generative modeling, the reconstructed probability distribution is represented
approximately by a graphical model or neural network -- the weights and biases serving as a parameterization of $P(\bm{x})$.  
After training, these generative models can be used to estimate the likelihood, or to produce samples, of new $\bm{x}$ in a way that generalizes and scales well.

This procedure can be extended to data produced by quantum devices, with the goal of reconstructing the quantum wavefunction (a complex generalization of a classical probability distribution).  NISQ devices with single-site control are particularly suited to this data-driven approach, since they can produce projective measurements of the state of individual qubits.  If a sufficient type and number of projective measurements can be obtained, industry-standard algorithms for
unsupervised learning of the relevant probability distributions (produced according to the Born rule) can be used to reconstruct the underlying quantum state.

Such data-driven state {\it reconstruction} may play by different rules than Hamiltonian-driven discovery of quantum states. In fact, the latter consists of obtaining a quantum state underlying a microscopic model (i.e.~a Hamiltonian), and it is a benchmark for quantum supremacy. Instead, the former assumes no knowledge of the Hamiltonian, but requires informationally-complete sets of measurement data on the quantum state.  The question of how efficiently the data-driven approach scales for wavefunctions of various structures of interest to physicists, and how it compares to the more conventional Hamiltonian-driven approach, is largely unanswered.

The most obvious role for a quantum state reconstructed via generative modeling is to produce new physical observables.  To be useful, this must be done in a tractable way that scales efficiently with increasing number of qubits, while generalizing well to unseen data.
The observables in question may be inaccessible to the device, such as those encoded in a basis for which no projective measurement was taken, or those (such as Renyi entanglement entropies) 
that require elaborate technical setups \cite{Islam15}.  Generative models are also capable of mitigating noise in the state preparation and measurement, a ubiquitous and defining condition in NISQ devices.  Finally, the ability to off-load the production of various observables to a parameterized model frees experimentalists 
to focus solely on the production of high-quality projective measurements.  It is this type of inelegant compromise that will allow machine learning techniques to contribute to the verification of quantum devices as they grow into the NISQ era and beyond.

In this paper, we review the development of generative modeling for quantum state reconstruction.  Beginning with the classical treatment of probability distributions, we motivate the use of a restricted Boltzmann machine (RBM), and demonstrate its ability to parameterize the thermal distribution of data drawn from a classical Ising model.  The same type of RBM is shown to faithfully reconstruct a real-positive wavefunction, and we demonstrate the production of non-trivial observables from the parameterized model. We then discuss extensions of standard RBMs to reconstruct complex wavefunctions and density matrices.  We end this review with a discussion of recent efforts in reconstruction of a real-world Rydberg atom quantum simulator.  Despite challenges in noisy state preparation and measurement, the demonstration of real experimental state reconstruction is a milestone for the use of machine learning in the NISQ era.

\section{Generative modeling}
Let us begin by considering an unknown probability distribution $P(\bm{x})$ defined over the $2^N$-dimensional space of binary states $\bm{x}=(x_1,\dots,x_N)$, and a set of data $\mathcal{D}=\{\bm{x}_k\}$ distributed according to $P(\bm{x})$. Can we infer the features of such distribution, such as regularities and correlations, directly from the observation of the data? In other words, can we discover an approximate representation $p(\bm{x})\approx P(\bm{x})$ from the limited-size dataset $\mathcal{D}$? The simplest approach consists of approximating the unknown probability with the frequency distribution obtained by inverting the measurement counts in the dataset:
\begin{equation}
p(\bm{x})=P_{\text{data}}(\bm{x})=\frac{1}{\|\mathcal{D}\|}\sum_{\bm{x}_k\in\mathcal{D}}\delta_{\bm{x},\bm{x}_k}\:.
\end{equation}
The validity of this approximation depends on the size of the system $N$ the entropy of the distribution and the size $\|\mathcal{D}\|$ of the dataset. For most practical purposes however, it fails to generalize the features of $P(\bm{x})$ beyond the training set. In contrast, generative modeling aims to discover an approximation of the unknown distribution that captures the underlying structure and it is also capable of generalization.

The first ingredient is a compact representation of the probability distribution, i.e.~a parametrization $p_{\bm{\lambda}}(\bm{x})$ in terms of a set of parameters $\bm{\lambda}$ whose number is much smaller than the size of the configuration space. Then, generative modeling consists of finding an  optimal set of parameters $\bm{\lambda}^*$ such that the parametric distribution $p_{\bm{\lambda}^*}(\bm{x})$ mimics the unknown distribution $P(\bm{x})$ underlying the finite number of dataset samples. In practice, this search is carried out through an optimization procedure, where the distance between the two distributions is minimized with respect to the model parameters $\bm{\lambda}$. The distance between two probability distributions can be quantified by the Kullbach-Leibler (KL) divergence
\begin{equation}
\mathbb{KL}_{\bm{\lambda}}(P\,\|\,p)=\sum_{\bm{x}}P(\bm{x})\log\frac{P(\bm{x})}{p_{\bm{\lambda}}(\bm{x})}\:,
\end{equation}
a non-symmetric statistical measure such that $\mathbb{KL}_{\bm{\lambda}}(P\,\|\,p_{\bm{\lambda}})>0$ and $\mathbb{KL}_{\bm{\lambda}}(P\,\|\,p_{\bm{\lambda}})=0$ if and only if $P=p_{\bm{\lambda}}$. By approximating the KL divergence with the measurement data, we obtain
\begin{equation}
\mathbb{KL}_{\bm{\lambda}}(P\,\|\,p)\approx\
-\frac{1}{\|\mathcal{D}\|}\sum_{\bm{x}\in\mathcal{D}}\log p_{\bm{\lambda}}(\bm{x})-\mathbb{H}_{\mathcal{D}}\:,
\label{Eq::KL}
\end{equation}
where $\mathbb{H}_{\mathcal{D}}$ is the dataset entropy. This quantity can be minimized iteratively by one of the many variants of the gradient descent algorithm. This procedure allows one to obtain a representation of the unknown distribution and generate new configurations that were not encountered in the learning stage. The most successful approach relies on the representation of $p_{\bm{\lambda}}(\bm{x})$ in terms of networks of artificial neurons.

\subsection{Artificial neural networks}
Artificial neural networks, the bedrock of modern machine learning and artificial intelligence~\cite{LeCun2015}, have a history spanning decades. Initially investigated to understand the process of human cognition, neural networks models are based on the idea that information (in the brain) has a distributed representation over a large collection of elementary units (neurons), and information processing occurs through the mutual interaction between neurons~\cite{PDP}. The fundamental ingredients are: i) a set of neurons, each one applying a simple type of computation to the input signal it receives; ii) a set of interactions defined over a graph structure connecting the neurons; iii) an external environment providing a ``teaching signal''; iv) a learning rule, i.e.~a prescription for modifying the interactions according to the external environment. 

The first artificial neuron capable of computation, the {\it perceptron}, was proposed by Frank Rosenblatt as early as 1957~\cite{Rosenblatt58}.  Based on the previous work of McCullogh and Pitts~\cite{McCullogh43}, the perceptron was capable of discriminating different classes of input patterns, a process called {\it supervised learning}. It was later shown that a single layer perceptron is only capable of learning linearly separable functions~\cite{Min69}, and since no learning algorithms were known for multi-layer perceptrons, the model was abandoned, leading to a decrease in both popularity and fundings of neural networks (called the first AI winter). The first resurgence of the field took place more than a decade later, with the invention of the backpropagation algorithm~\cite{HintonBackProp} and the Boltzmann machine (BM)~\cite{Ackley85}. The latter, was directly built on the connection between cognitive science and statistical mechanics, established by the works of condensed matter physicists William Little~\cite{Little74,Little78} and John Hopfield~\cite{Hopfield82}.

\subsubsection{The Hopfield model}
The Hopfield network, introduced in 1982 as a model for associative memories~\cite{Hopfield82}, was inspired by the concept of emergence in condensed matter physics, where complex behaviors effectively emerge from the mutual interactions of a large number of degrees of freedom. 
In this context, Hopfield formulated a physics-inspired model of cognition for the task of recovering a corrupted memory. By regarding a memory as a state $\bm{x}$ containing $N$ bits of information, the corresponding network consists of $N$ binary neurons fully connected with symmetric weights (or interactions), described by an energy function
\begin{equation}
E(\bm{x})=-\sum_{ij}W_{ij}x_ix_j\:.
\label{Eq::Hopfield_energy}
\end{equation}
Each neuron in the network carries out the computation of $\sum_jW_{ij}x_j$, and update itself according to the following rule:
\begin{equation}
x_i = \begin{cases} 
1, & \mbox{if } \sum_jW_{ij}x_j>0\\ 
0, & \mbox{otherwise.} \end{cases}
\label{hop_net_out}
\end{equation}
Since the energy difference between the two possible states of the $i$-th neuron is $\Delta E_i=\sum_jW_{ij}x_j$, the dynamics resulting from the  asynchronous update of the neurons monotonically minimizes the total energy. Therefore, given an initial state, the network evolves in time by following the above equation of motion until a stable configuration (i.e.~a local energy minimum) is found. In the context of associative memories, given a set of desired memory states $\{\bar{\bm{x}}_1,\bar{\bm{x}}_2,\dots\}$, there exists a learning rule to modify the interactions $\bm{W}$ in such a way that these states become local minima in the energy landscape~\cite{Hopfield82}. Thus, if the network is initialized to a corrupted memory $\bar{\bm{x}}_k+\bm{\delta}$ which is sufficiently close to the true state (i.e. small $\bm{\delta}$), the network is able to recover the correct memory simply by evolving with its equations of motion.

\subsubsection{The Boltzmann machine}
The two major limitations of the Hopfield model are the tendency for the network to get trapped into local minima, and its restricted capacity. Nevertheless, it suggested the important connection between cognitive science and statistical physics, which was was further strengthened with the invention of the BM by Ackley, Hinton and Sejnowski in 1985~\cite{Ackley85}. Similarly to the Hopfield model, the BM consists of a set of $N$ binary neurons interacting with energy given in Eq.~\ref{Eq::Hopfield_energy}. However, in order to allow the system to escape local minima, the neural network is placed at thermal equilibrium at some inverse temperature $\beta=1/T$. The update rule becomes now stochastic, with the $i$-th neuron activating ($x_i=1$) with probability
\begin{equation}
p_i = \frac{1}{1+e^{-\beta\Delta E_i}}\:,
\label{BM_update}
\end{equation}
where $\Delta E_i$ is once again the energy difference between its two internal states. As the temperature goes to zero ($\beta\rightarrow\infty)$, one recovers the Hopfield model with a deterministic dynamics minimizing energy. Instead, for a stochastic dynamics at finite temperature, the network minimizes the free energy instead, equilibrating to the canonical Boltzmann distribution:
\begin{equation}
p(\bm{x})=\frac{1}{Z}e^{-\beta E(\bm{x})}\qquad,\qquad Z=\sum_{\bm{x}}e^{-\beta E(\bm{x})}.
\end{equation}

The BM is one of the simplest examples of generative models. In fact, the set of interactions can be considered as tunable parameters, resulting into a parametric distribution $p_{\bm{\lambda}}(\bm{x})$ (with $\bm{\lambda}=\bm{W}$). Then, the interactions can be modified following an unsupervised learning procedure in order for the network distribution to mimic an unknown probability distribution underlying a given set of data points $\mathcal{D}=\{\bm{x}\}$. By minimizing the statistical divergence between the data and model distribution (Eq.~\ref{Eq::KL}), one obtains the following learning rule for the parameters~\cite{Ackley85},
\begin{equation}
\Delta W_{ij} = \beta\Big[\langle x_ix_j\rangle_{\mathcal{D}}-\langle x_ix_j\rangle_{p(\bm{x})}\Big]\:.
\end{equation}
In the {\it positive} phase, the weight $W_{ij}$ is increased according to the average value of $x_ix_j$ over the data points in $\mathcal{D}$, corresponding to a traditional Hebbian learning (i.e.~``neurons that fire together wire together''). This term effectively lowers the energy of all configurations that are compatible with the dataset, thus increasing their probability. In contrast, in the {\it negative} phase, the same process occurs with the reverse sign, decreasing the probability of configurations generated by the BM when running freely at thermal equilibrium. Clearly, as the two averages coincide, the BM distribution reproduces the dataset and there is not net change in the parameter. Otherwise, the network is trying to {\it unlearn} configurations generated at equilibrium that lead to an imbalance with respect to the data. It is interesting to note how this {\it learning and unlearning} process had been already proposed in an ad-hoc way by Hopfield to eliminate spurious minima in his model of associate memories~\cite{Hopfield83}.

The major limitation of this network is the structure of the energy function, allowing the BM to capture only pairwise correlation in the data (e.g. it cannot learn the XOR function~\cite{Hinton:1986:LRB}). The simplest way to increase the reach of its representational capabilities is to introduce an auxiliary set of neurons which do not appear in the input space of the data. The full network is then divided as $\bm{x}=(\bm{v},\bm{h})$, where $\bm{v}$ are called {\it visible units}, corresponding to the degrees of freedom in the dataset, and $\bm{h}$ are called {\it hidden units}~\cite{Ackley85}. In order to derive a learning rule for the network parameters, one needs to eliminate the hidden degrees of freedom so that an explicit distribution over the visible neurons $p_{\bm{\lambda}}(\bm{v})=\sum_{\bm{h}}p_{\bm{\lambda}}(\bm{v},\bm{h})$ can be obtained. Therefore, to attain a tractable marginal distribution, one can restrict the interactions between neurons on different layers, resulting in the famous {\it restricted Boltzmann machine} (RBM).

\begin{figure}[t]
\includegraphics[width=\columnwidth]{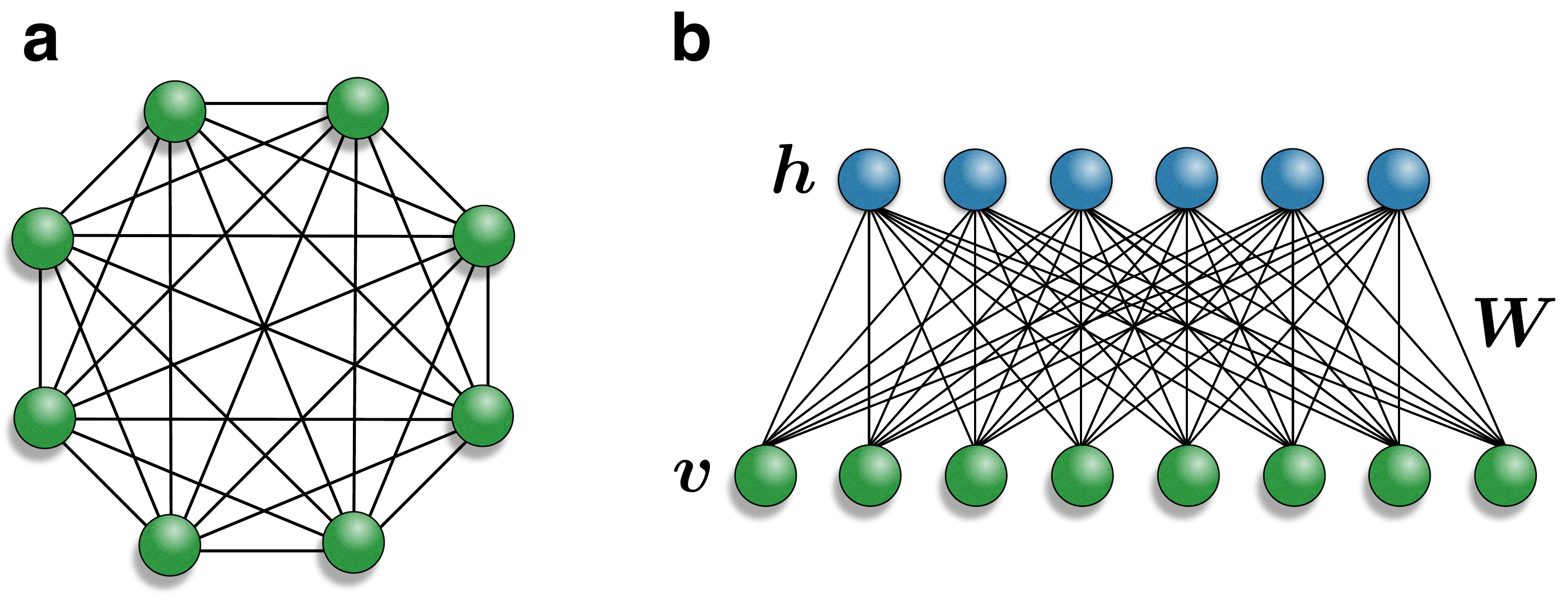}
\caption{Probabilistic graphical models. ({\it a}) A fully connected neural network, which can represent both the Hopfield model or Boltzmann machine, depending on the update rule. ({\it b}) A restricted Boltzmann machine, with a set of symmetric weights $\bm{W}$ connecting the visible and the hidden layer.}
\label{Fig::rbm}
\end{figure}

\subsection{Restricted Boltzmann machines}
The RBM, originally introduced by Smolensky under the name of Harmonium~\cite{Smolensky1986}, is a probabilistic graphical model with energy
\begin{equation}
E_{\bm{\lambda}}(\bm{v},\bm{h})=-\sum_{ij}W_{ij}h_iv_j-\sum_jb_jv_j-\sum_ic_ih_i\:,
\end{equation}
where we have added bias terms (i.e.~magnetic fields) $\bm{b}$ and $\bm{c}$ for the visible and hidden layer respectively. The set of tunable parameters is now $\bm{\lambda}=(\bm{W},\bm{b},\bm{c})$ (Fig.~\ref{Fig::rbm}b). The marginal distribution, obtained by tracing out the hidden neurons, can be calculated analytically 
\begin{equation}
p_{\bm{\lambda}}(\bm{v})=\sum_{\bm{h}}p_{\bm{\lambda}}(\bm{v},\bm{h})=
\frac{1}{Z_{\bm{\lambda}}}\sum_{\bm{h}}e^{-E_{\bm{\lambda}}(\bm{v},\bm{h})}=
\frac{1}{Z_{\bm{\lambda}}}e^{\:-\mathcal{E}_{\bm{\lambda}}(\bm{v})}\:,
\end{equation}
where we set the inverse temperature to $\beta=1$ and we introduced the new energy function
\begin{equation}
\mathcal{E}_{\bm{\lambda}}(\bm{v})=-\sum_jb_jv_j-\sum_i\log\left(1+e^{\:\sum_{j}W_{ij}v_j+c_i}\right)\:.
\end{equation}
The energy $\mathcal{E}_{\bm{\lambda}}(\bm{v})$ defines an effective system consisting of the visible neurons only. We can see that the energy contains two terms: a mean-field contribution, proportional to the visible bias $\bm{b}$ and a non-linearity containing correlations between visible neurons at all orders. The particular structure of such an effective energy allows the RBM distribution $p_{\bm{\lambda}}(\bm{v})$ to be a universal function approximator of discrete distributions~\cite{LeRoux08}. This means that, given a large enough number of hidden neurons, any function of discrete binary variables can be approximated to arbitrary precision. However, in the worst case scenario, the number of hidden neurons may grow exponentially with the visible layer.

\subsubsection{Unsupervised learning}
The goal of unsupervised learning is to discover a set of parameters so that the RBM distribution mimics the unknown distribution underlying a dataset $\mathcal{D}=\{\bm{v}_1,\bm{v}_2,\dots\}$ of visible samples. The cost function, given by the KL divergence from Eq.~\ref{Eq::KL}, is
\begin{equation}
\begin{split}
\mathcal{C}_{\bm{\lambda}}&=-\frac{1}{\|\mathcal{D}\|}\sum_{\bm{v}\in\mathcal{D}}\log p_{\bm{\lambda}}(\bm{v})=-\frac{1}{\|\mathcal{D}\|}\sum_{\bm{v}\in\mathcal{D}}\mathcal{E}_{\bm{\lambda}}(\bm{v})-\log Z_{\bm{\lambda}}\:,
\end{split}
\end{equation}
where we have omitted the constant entropy term $\mathbb{H}_{\mathcal{D}}$. The learning rule for the RBM parameters is obtained by taking the gradient of $\mathcal{C}_{\bm{\lambda}}$
\begin{equation}
\Delta\bm{\lambda}\propto-\nabla_{\bm{\lambda}}\mathcal{C}_{\bm{\lambda}}=\big\langle\nabla_{\bm{\lambda}}\mathcal{E}_{\bm{\lambda}}(\bm{v})\big\rangle_{\mathcal{D}}-\big\langle\nabla_{\bm{\lambda}}\mathcal{E}_{\bm{\lambda}}(\bm{v})\big\rangle_{p_{\bm{\lambda}}(\bm{v})}\:
\label{Eq::rbm_grad}
\end{equation}
where the gradients $\nabla_{\bm{\lambda}}\mathcal{E}_{\bm{\lambda}}(\bm{v})$ are straightforwards to calculate. Similar to the regular BM, the gradient contains two competing terms, driven respectively by the data and the RBM distribution. The first term (the positive phase) is trivial to compute, being an average over the data:
\begin{equation}
\big\langle\nabla_{\bm{\lambda}}\mathcal{E}_{\bm{\lambda}}(\bm{v})\big\rangle_{\mathcal{D}}=\frac{1}{\|\mathcal{D}\|}\sum_{\bm{v}\in\mathcal{D}}\nabla_{\bm{\lambda}}\mathcal{E}_{\bm{\lambda}}(\bm{v})\:.
\end{equation}
Conversely, the calculation of the negative phase is in general intractable. It needs to be approximated using a Monte Carlo simulation,
\begin{equation}
\begin{split}
\big\langle\nabla_{\bm{\lambda}}\mathcal{E}_{\bm{\lambda}}(\bm{v})\big\rangle_{p_{\bm{\lambda}}(\bm{v})}&=\frac{1}{Z_{\bm{\lambda}}}\sum_{\bm{v}}p_{\bm{\lambda}}(\bm{v})\nabla_{\bm{\lambda}}\mathcal{E}_{\bm{\lambda}}(\bm{v})\\
&\approx\frac{1}{M}\sum_{k=1}^M\nabla_{\bm{\lambda}}\mathcal{E}_{\bm{\lambda}}(\bm{v}_k)\:,
\label{Eq::negative_phase}
\end{split}
\end{equation}
where the configurations $\bm{v}_k$ are drawn from a Markov chain running on the distribution $p_{\bm{\lambda}}(\bm{v})$.

The sampling stage to estimate the negative phase, which is the computational bottleneck of the training, is aided by the restricted nature of the RBM graph. In fact, because of that, neurons in a given layer are conditionally independent of one another. That is, due to the lack of intra-layer connections, the conditional probabilities for the neurons in one layer conditioned on the current state of the other factorize over each individual neuron,
\begin{equation}
p_{\bm{\lambda}}(\bm{v}\:|\:\bm{h})=\prod_jp_{\bm{\lambda}}(v_j\:|\:\bm{h})\quad,\quad p_{\bm{\lambda}}(\bm{h}\:|\:\bm{v})=\prod_i p_{\bm{\lambda}}(h_i\:|\:\bm{v})\:,
\end{equation}
and can be easily calculated analytically~\cite{igel_rbm}. When running the Markov chain to collect the statistic in Eq.~\ref{Eq::negative_phase}, one can sample the state of each neuron in one layer simultaneously using the above conditional probabilities,  alternating between visible and hidden layer. This sampling strategy is called {\it block Gibbs sampling}.

\subsubsection{Training by contrastive divergence}
The calculation of the negative phase, even if carried out using block Gibbs sampling, is still computationally intensive. In fact, at each training iteration, the RBM needs to reach its equilibrium distribution $p_{\bm{\lambda}}(\bm{v})$ before collecting the statistics for the negative phase calculation. Furthermore, the gradient in Eq.~\ref{Eq::rbm_grad} can display a large variance, being the difference of two averages computed from two different distributions. A solution to both these issues is to consider a different cost function.  Namely, the {\it contrastive divergence} (CD) between the data and the RBM after a sequence of $k$ block Gibbs sampling steps is~\cite{Hinton02},
\begin{equation}
\mathbb{CD}_k=\mathbb{KL}(P_{\text{data}}\:|\:p_{\bm{\lambda}})-\mathbb{KL}(p^{(k)}_{\bm{\lambda}}\:|\:p_{\bm{\lambda}})\:,
\end{equation}
where $p^{(k)}_{\bm{\lambda}}$ is the probability distribution of the visible layer after $k$ steps. The new update from the gradient of $\mathbb{CD}_k$ becomes~\cite{Bengio09,igel_cd,Carreira}
\begin{equation}
\Delta\bm{\lambda}\propto\big\langle\nabla_{\bm{\lambda}}\mathcal{E}_{\bm{\lambda}}(\bm{v})\big\rangle_{\mathcal{D}}-\big\langle\nabla_{\bm{\lambda}}\mathcal{E}_{\bm{\lambda}}(\bm{v})\big\rangle_{p^{(k)}_{\bm{\lambda}}(\bm{v})}\:.
\end{equation}
The resulting CD training consists of initializing the RBM to a random sample from the dataset $\mathcal{D}$ and using the visible state after $k$ steps of block Gibbs sampling to evaluate the negative phase.

Once the gradient of the cost function is calculated, the parameters $\bm{\lambda}$ are updated with a gradient descent algorithm. The simplest one, called {\it stochastic gradient descent}, uses a random set of data to evaluate the positive phase and performs the update $\Delta\bm{\lambda}=-\eta\nabla_{\bm{\lambda}}\mathcal{C}_{\bm{\lambda}}$ where $\eta$ is the step-size of the update, also called the {\it learning rate}. The total number of data samples used for the update is called the {\it batch size}. Other algorithms can be used to speed up the convergence~\cite{momentum} and tune the learning rate in an adaptive way~\cite{adadelta,adam}. Furthermore, an additional term should be added to the cost function to help generalization, i.e. avoid the overfitting of data points. A common choice is {\it weight decay}~\cite{NIPS1991_563} regularization, which penalizes large value of the weights. We refer the reader to Ref~\cite{Hinton2012} for more details on the practical training of RBMs and a description of the various training hyperparameters (and how to choose them).

\section{Quantum state reconstruction}
Image that an experimental NISQ apparatus in the laboratory containing $N$ qubits is prepared in some quantum state of interest, described by a density operator $\hat{\varrho}$. Because of the practical limitations imposed by the hardware, measurements of properties of interest might be costly, or not technically possible. It is then highly desirable to be able to reconstruct the quantum state $\hat{\varrho}$ from simple, experimentally feasible measurements.

The traditional approach for reconstructing a quantum state from measurement data is called {\it quantum state tomography} (QST)~\cite{vogel89,Jezek2003,Banaszek2013}. 
A typical procedure consists of maximum-likelihood reconstruction of a density operator parametrized as $\hat{\rho}\propto\hat{T}^\dagger\hat{T}$~\cite{James01}, where $\hat{T}$ is a tri-diagonal hermitian matrix, enforcing the positive semi-definite requirement on $\hat{\rho}$. 
Such procedures assume no {\it a priori} phase structure to the quantum state, or even whether it is necessarily pure.  Such ``full'' QST therefore typically scales exponentially.
Given this, full QST can only be effectively carried out for systems with a relatively small number of particles or qubits~\cite{full_qst}. In general however, physical quantum states -- such as ground states of local Hamiltonians -- possess large degrees of structure.  This often makes it possible to obtain a compact representation with resources scaling polynomially with the size of the Hilbert space. The most notable example are matrix product states (MPS), which have been used to successfully reconstruct quantum states outside the reach of full QST~\cite{Cramer2010,Lanyon2017}. However, so-called MPS tomography inherits the intrinsic limitations of the MPS representation, namely the restriction to one dimensional systems and low-entangled states, which limits the reconstruction of short-time dynamics, for example. 
The inherent structure of a quantum state can also be exploited in alternative ways, such as in permutationally invariant QST~\cite{toth2010,Moroder_2012} or compressed sensing~\cite{Gross2010}.

In this section, we overview a machine learning-based approach to QST, and show that unsupervised learning of generative models provides a very natural framework for reconstructing quantum many-body states. 
As described in the last section, RBMs offer a generative modeling framework that is conceptually interpretable in the context of statistical physics.  In addition, they have been more widely explored in applications in classical and quantum state reconstruction than any other generative model. We start by considering the simplest case of reconstructing a thermal state in the classical limit, and proceed with increasing complexity to the case of pure quantum wavefunctions and finally density operators.

\subsection{Classical limit}
We start with the reconstruction of a physical system at thermal equilibrium, and consider the classical limit where the Hamiltonian under consideration is diagonal in the measurement basis $\{|\bm{\sigma}\rangle\}$ . The density operator we aim to reconstruct simply reduces to
\begin{equation}
\hat{\varrho}=\frac{e^{-\beta\hat{H}}}{\text{Tr}[e^{-\beta\hat{H}}]}=
\sum_{\bm{\sigma}}P_{\beta}(\bm{\sigma})|\bm{\sigma}\rangle\langle\bm{\sigma}|
\end{equation}
where $P_{\beta}(\bm{\sigma})=e^{-\beta H(\bm{\sigma}^z)}/Z_\beta$ is the classical Boltzmann distribution in the canonical ensemble and $Z_\beta$ its partition function. State reconstruction is inherently a classical problem here, corresponding to the unsupervised learning of the distribution $P_{\beta}(\bm{\sigma})$. A simple yet non-trivial example is given by the Ising model, where $N$ spins interact with Hamiltonian
\begin{equation}
H(\bm{\sigma})=-\sum_{\langle ij\rangle}\sigma_i\sigma_j
\end{equation}
with the sum running over nearest neighbours on a lattice. In two dimensions, the spin system displays ferromagnetic order at low temperature and a high-temperature disordered state, separated by a continuous phase transition.

\begin{figure}[t]
\includegraphics[width=0.79\columnwidth]{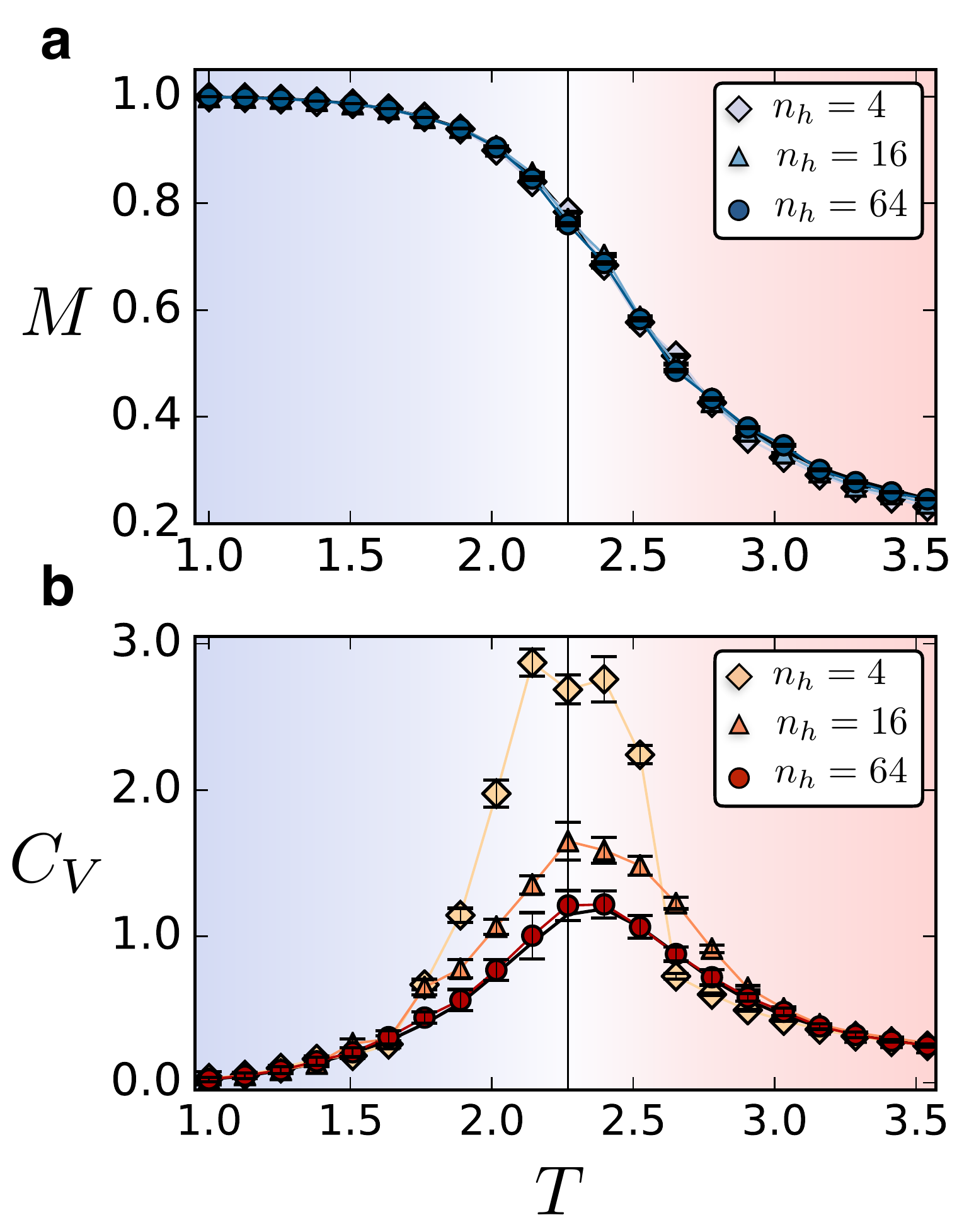}
\caption{Learning the thermodynamics of the classical Ising model at thermal equilibrium. Comparison of the average values of the magnetization ({\it a}) and the specific heat ({\it b}) between the exact values calculated on the dataset (sampled by MC) and the values generated after the reconstruction, for an increasing number of hidden neurons in the RBM. Figure reproduced from reference~\cite{TorlaiThesis}.}
\label{Fig::ising}
\end{figure}

As first demonstrated in Ref.~\cite{torlai_learning_2016}, different RBMs can be trained on datasets containing spin configurations at different temperatures across the phase diagram, generated by importance sampling the partition functions using Monte Carlo simulations~\cite{metropolis}. The quality of the reconstruction can be assessed by comparing expectation values of thermodynamics observables generated by the RBM with the exact values calculated on the datasets. In Fig.~\ref{Fig::ising} we report such a comparison for the magnetization and specific heat, with a varying number of hidden units in the RBM. While the magnetization converges very quickly -- since it it explicitly encoded in the dataset -- a larger number of hidden units is required to accurately reproduce the specific heat, particularly in the presence of large fluctuations at the critical point. Finally, we point out the curious observation that the quality of the reconstruction does not obviously improve for deep versions of the RBMs~\cite{Morningstar17}, such as deep belief networks~\cite{hinton_dbn} or deep Boltzmann machines~\cite{dbm}.

\subsection{Positive wavefunctions}
We now turn to quantum states described by pure density operators $\hat{\varrho}=|\Psi\rangle\langle\Psi|$, where the wavefunction has representation $|\Psi\rangle=\sum_{\bm{\sigma}}\Psi(\bm{\sigma})|\bm{\sigma}\rangle$ with coefficients  $\Psi(\bm{\sigma})=\langle\bm{\sigma}|\Psi\rangle$ in the measurement basis $\{|\bm{\sigma}\rangle\}$. In addition, we assume for now that the pure state $|\Psi\rangle$ has a real and positive representation in this basis, $\Psi(\bm{\sigma})\in\mathbb{R}$ and $\Psi(\bm{\sigma})>0\:\:\forall|\bm{\sigma}\rangle$. Under this assumption, valid for example for ground states of so-called ``stoquastic" Hamiltonians~\cite{Bravyi2008}, the wavefunction $|\Psi\rangle$ is uniquely characterized by the probability distribution underlying a set of projective measurements, given by the Born rule $P(\bm{\sigma})=|\Psi(\bm{\sigma})|^2$. The inherently probabilistic nature of quantum mechanics provides a simple and natural way to define a representation of a pure and positive quantum state in terms of an RBM~\cite{torlai_Tomo},
\begin{equation}
\psi_{\bm{\lambda}}(\bm{\sigma})=\sqrt{p_{\bm{\lambda}}(\bm{\sigma})}=
\frac{1}{\sqrt{Z_{\bm{\lambda}}}}e^{-\mathcal{E}_{\bm{\lambda}}(\bm{\sigma})/2}\:.
\end{equation}
Note that, since RBMs are universal approximators of any discrete probability distribution, provided the number of hidden units in the network is sufficiently large, the RBM wavefunction $\psi_{\bm{\lambda}}(\bm{\sigma})$ is capable of representing any positive quantum state to arbitrary accuracy.

Because of the positivity of the target state, quantum state reconstruction in this case is equivalent to conventional RBM unsupervised learning. Upon minimization of the KL divergence between the projective measurement distribution and the RBM distribution,
\begin{equation}
\begin{split}
\mathcal{C}_{\bm{\lambda}}&=\sum_{\bm{\sigma}}|\Psi(\bm{\sigma})|^2\log\frac{|\Psi(\bm{\sigma})|^2}{|\psi_{\bm{\lambda}}(\bm{\sigma})|^2}\\
&=-\frac{1}{\|\mathcal{D}\|}\sum_{\bm{\sigma}_k\in\mathcal{D}}\log p_{\bm{\lambda}}(\bm{\sigma})-\mathbb{H}_{\mathcal{D}},
\end{split}
\end{equation}
the RBM wavefunction approximates the target state $\psi_{\bm{\lambda}}\sim\Psi$ as desired.

\subsubsection{Measurement of physical observables} 
By discovering a set of parameters that successfully minimizes the cost function, the RBM builds an internal representation of the unknown target wavefunction and can be sampled to compute expectation values of physical observables. If the observable $\hat{\mathcal{O}}$ is diagonal in the measurement basis, its expectation value reduces to a thermal average with respect to the RBM distribution
\begin{equation}
\langle\hat{\mathcal{O}}\rangle=\langle\psi_{\bm{\lambda}}|\hat{\mathcal{O}}|\psi_{\bm{\lambda}}\rangle=\sum_{\bm{\sigma}}p_{\bm{\lambda}}(\bm{\sigma})\mathcal{O}_{\bm{\sigma\sigma}}\:,
\end{equation}
which can be approximated by a Monte Carlo average using block Gibbs sampling. Calculations of diagonal observables provide a direct verification of the quality of the training, since the expectation values can be compared with those calculated directly on the training dataset. 

More interestingly, the RBM allows one to estimate average values of observables which are off-diagonal in the measurement basis. For this case, the expectation value reduces to the average $\langle\hat{\mathcal{O}}\rangle=\langle\mathcal{O}_L(\bm{\sigma})\rangle_{p_{\bm{\lambda}}(\bm{\sigma})}$, where 
\begin{equation}
\mathcal{O}_L(\bm{\sigma})=\sum_{\bm{\sigma}^\prime}\frac{\psi_{\bm{\lambda}}(\bm{\sigma}^\prime)}{\psi_{\bm{\lambda}}(\bm{\sigma})}\mathcal{O}_{\bm{\sigma}^\prime\bm{\sigma}},
\end{equation}
is the so-called {\it local estimate} of the observables~\cite{becca_sorella_2017}. Provided the matrix representation of $\hat{\mathcal{O}}$ is sufficiently sparse in the measurement basis (i.e.~the number of off-diagonal elements that are non-zero scales polynomially with $N$), its expectation value can be efficiently estimated with Monte Carlo.

Another important quantity amenable to calculation with an RBM is the entanglement of a subsystem $A$, which for pure states is quantified by the Renyi entropy~\cite{renyi1961}
\begin{equation}
S_\alpha(\hat{\rho}_A)=\frac{1}{1-\alpha}\log\text{Tr}(\hat{\rho}_A^\alpha)\:,
\end{equation}
with $\hat{\rho}_A=\text{Tr}_{A^\perp}|\Psi\rangle\langle\Psi|$ the reduced density matrix of $A$. For the case of $\alpha=2$, the entanglement entropy can be calculated by considering two identical replicas of the original system, and computing the overlap between the states with and without the configurations of subregion $A$ swapped between the replicas
\begin{equation}
\begin{split}
S_2(\hat{\rho}_A)&=-\log\text{Tr}(\hat{\rho}_A^2)\\
&=-\log\Big[\langle\Psi_1|\otimes\langle\Psi_2|\text{Swap}_A|\Psi_1\rangle\otimes|\Psi_2\rangle\Big]\:.
\end{split}
\end{equation}
Calculations of the entanglement entropy via this procedure has been successfully carried out in numerical simulations using different flavors of quantum Monte Carlo algorithms~\cite{hastings10,Kallin11, VMC_swap,St_phan_2013}. In the experimental context, the measurement of the swap operator has been performed for ultracold bosons in optical lattices by means of quantum interference~\cite{Islam15}. Interestingly, the above implies that same measurement can be efficiently implemented using the RBM wavefunction, where instead of replicating the 
experimental NISQ hardware in the laboratory, one can first train the RBM to learn the experimental wavefunction, and then perform the calculation of the swap operator by replicating the neural network~\cite{torlai_Tomo}.

\begin{figure}[t]
\includegraphics[width=0.79\columnwidth]{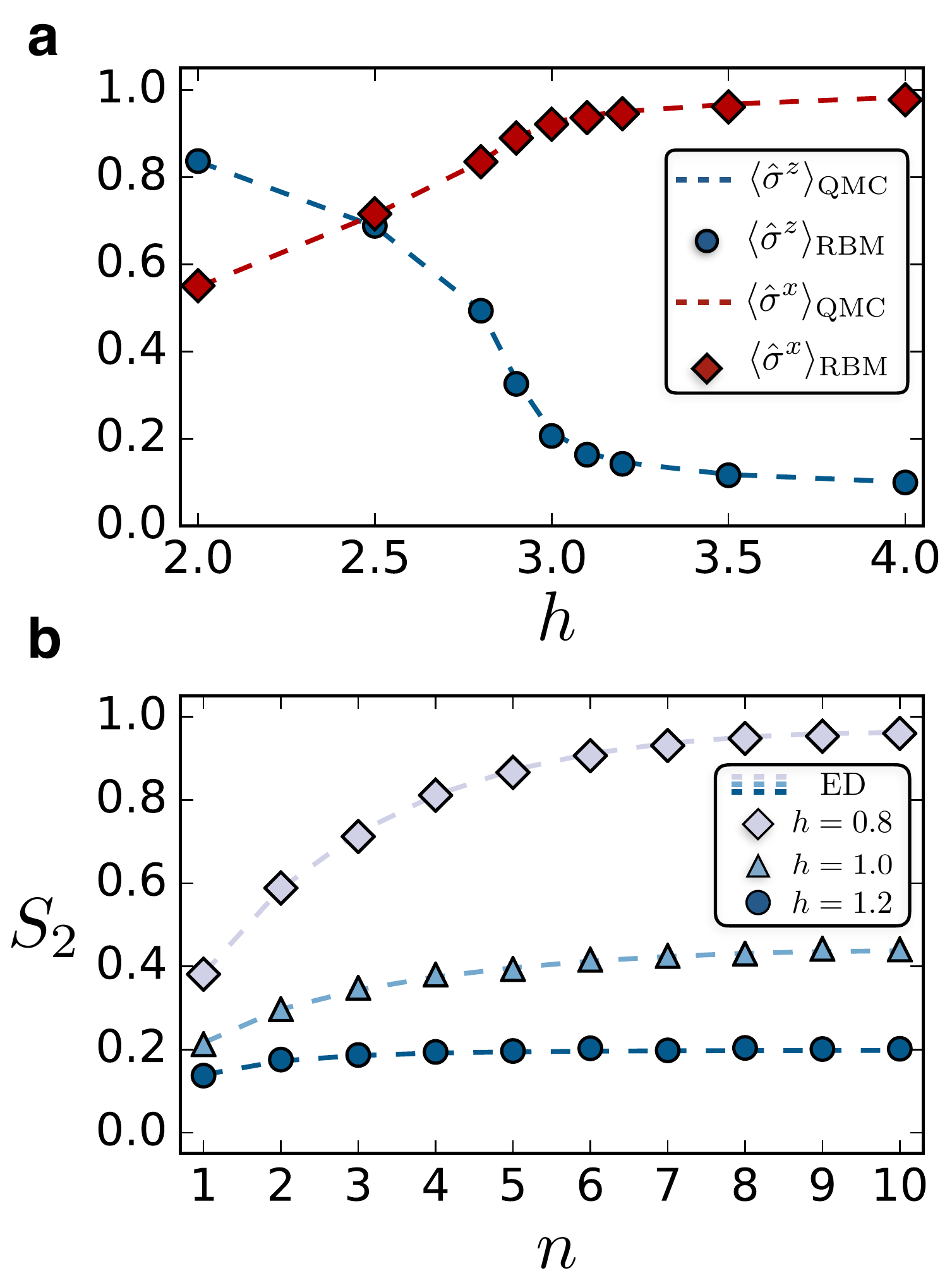}
\caption{Quantum state reconstruction of the transverse-field Ising model. ({\it a}) Longitudinal and transverse magnetization for the Ising model on a square lattice with $N=144$ and open boundaries as a function of the transverse field $h$. ({\it b}) Entanglement entropy for an open chain with $N=20$ spins as a function of the subsystem size, for transverse field below, above and at the quantum critical point. The datasets were generated with Quantum Monte Carlo ({\it a}) and exact diagonalization ({\it b}). Figure reproduced from~\cite{TorlaiThesis}.}
\label{Fig::tfim}
\end{figure}

\subsubsection{Reconstructing quantum spins on a lattice}
As an example, we review a numerical experiment for the quantum reconstruction of the ground state of the transverse-field Ising model, with Hamiltonian
\begin{equation}
\hat{H}=-\sum_{\langle ij\rangle}\hat{\sigma}^z_i\hat{\sigma}^z_j-h\sum_i\hat{\sigma}^x_i\:.
\end{equation}
This spin system undergoes a quantum phase transition between a ferromagnetic state for a small value of the transverse field $h$, and a paramagnetic state for large $h$. Measurement data in the $\{|\bm{\sigma}^z\rangle\}$ basis can be generated with standard methods \cite{white_dmrg,evertz}. 
Similar to the case of the classical Ising model above, different RBMs are trained at different values of the transverse field, and then sampled to generate expectation values of observables~\cite{torlai_Tomo}. Fig.~\ref{Fig::tfim} shows the reconstruction of the average diagonal and off-diagonal (transverse) magnetizations for the quantum Ising model 
on a square lattice, and the entanglement entropy for the one-dimensional chain, calculated using the swap operator between replicated copies of the neural network.

\subsection{Complex wavefunctions}
The assumption of a pure and positive quantum state enables RBM reconstruction with a favorable scaling with respect to the number of particles in the system. In general however, experimental quantum states might violate this assumption, containing a sign or a phase structure, where the coefficients of the wavefunction can be both positive and negative, or complex-valued $\Psi(\bm{\sigma})=|\Psi(\bm{\sigma})|e^{i\phi(\bm{\sigma})}$. A sign structure often appears in ground states of non-stoquastic Hamiltonians, such as quantum spins with competing interactions on frustrated lattices, or fermions. In this case, data from a single measurement basis is clearly not sufficient to fully  capture the quantum state, since the corresponding probability distribution $P(\bm{\sigma})=|\Psi(\bm{\sigma})|^2$ does not contain any fingerprints of the sign structure. Thus, reconstruction of the quantum state requires measurement in additional bases.

The first step for generalizing the RBM reconstruction to complex-valued wavefunctions is to define an appropriate neural-network parametrization of the quantum state. The most straightforward way consists of adding a phase factor to the positive RBM wavefunction defined in the previous section, $\psi_{\bm{\lambda\mu}}(\bm{\sigma})=\sqrt{p_{\bm{\lambda}}(\bm{\sigma})}e^{i\theta_{\bm{\mu}}(\bm{\sigma})}$. There is a large amount of freedom in choosing the nature of the phase function $\theta_{\bm{\mu}}(\bm{\sigma})$ in term of additional network parameters $\bm{\mu}$, and it needs not to be restricted to generative models. In fact, any feedforward neural network, such as convolutional networks~\cite{NIPS2012_4824}, could be used to this end. 
Another powerful way to adapt the RBM to quantum states is by using complex-valued weights and biases~\cite{Carleo}.
In this review we will use an additional RBM to capture the phases,
 leading to the following neural-network wavefunction~\cite{torlai_Tomo}
\begin{equation}
\psi_{\bm{\lambda\mu}}(\bm{\sigma})=\frac{1}{\sqrt{Z_{\bm{\lambda}}}}e^{-(\mathcal{E}_{\bm{\lambda}}(\bm{\sigma})+i\mathcal{E}_{\bm{\mu}}(\bm{\sigma}))/2}\:.
\end{equation}
Note that the generation of configurations in the reference basis corresponds to sampling the distribution $|\psi_{\bm{\lambda}}(\bm{\sigma})|^2=p_{\bm{\lambda}}(\bm{\sigma})$ which does not depend on the phases and can be then carried out by using block Gibbs sampling on the RBM with parameters $\bm{\lambda}$.

\subsubsection{Learning the phase structure} 
The reconstruction of a phase structure requires performing additional measurements in bases different than the reference one where the RBM wavefunction is expressed. This involves applying a unitary transformation $\hat{\mathcal{U}}$ to the quantum state,
\begin{equation}
\Psi(\bm{\sigma}^{\bm{b}})=\sum_{\bm{\sigma}}\mathcal{U}_{\bm{\sigma}^{\bm{b}}\bm{\sigma}}\Psi(\bm{\sigma})\:,
\end{equation}
where $|\bm{\sigma}^{\bm{b}}\rangle=|\sigma_1^{b_1},\dots,\sigma_N^{b_N}\rangle$ and $b_j$ identifies a particular choice of local basis for the $j$-th degree of freedom. The corresponding probability distribution after the measurement, $P(\bm{\sigma}^{\bm{b}})=|\Psi(\bm{\sigma}^{\bm{b}})|^2$, contains partial information on the phases and can be used to reconstruct the complex state. 
In general, such a unitary transformation consists of a collection of independent rotations of the local Hilbert spaces. The number and the type of rotations required to extract sufficient information to learn a phase depends on the structure of the specific quantum state under reconstruction.

Given a dataset $\mathcal{D}=\{\bm{\sigma}^{\bm{b}}\}$ of measurements in different bases, the RBM reconstruction can be realized by minimizing the total KL divergence in all bases,
\begin{equation}
\begin{split}
\mathcal{C}_{\bm{\lambda\mu}}&=-\frac{1}{\|\mathcal{D}\|}\sum_{\bm{\sigma^b}\in\mathcal{D}}\log|\psi_{\bm{\lambda\mu}}(\bm{\sigma}^{\bm{b}})|^2\\
&= - \frac{1}{\|\mathcal{D}\|}\sum_{\bm{\sigma^b}\in\mathcal{D}}\Bigg[\log\left(\sum_{\bm{\sigma}}\mathcal{U}_{\bm{\sigma}^{\bm{b}},\bm{\sigma}}{\psi}_{\bm{\lambda\mu}}(\bm{\sigma})\right)+c.c\Bigg]\:,
\end{split}
\end{equation}
where we have omitted the constant entropy term. By taking the gradients with respect to the parameters one obtains:
\begin{equation}
\nabla_{\bm{\lambda}}\mathcal{C}_{\bm{\lambda\mu}}=\frac{1}{\|\mathcal{D}\|}\sum_{\bm{\sigma^b}\in\mathcal{D}}\mathbb{R}\text{e}\bigg[\big\langle\nabla_{\bm{\lambda}}\mathcal{E}_{\bm{\lambda}}(\bm{\sigma})\big\rangle_{Q^{\bm{b}}(\bm{\sigma})}\bigg]-\big\langle\mathcal{E}_{\bm{\lambda}}(\bm{\sigma})\big\rangle_{p_{\bm{\lambda}}}\:,
\end{equation}
\begin{equation}
\nabla_{\bm{\mu}}\mathcal{C}_{\bm{\lambda\mu}}=-\frac{1}{\|\mathcal{D}\|}\sum_{\bm{\sigma^b}\in\mathcal{D}}\mathbb{I}\text{m}\bigg[\big\langle\nabla_{\bm{\mu}}\mathcal{E}_{\bm{\mu}}(\bm{\sigma})\big\rangle_{Q^{\bm{b}}(\bm{\sigma})}\bigg]\:,
\label{grad_mu}
\end{equation}
where the averages over the quasi-probability distribution $ Q^{\bm{b}}(\bm{\sigma})=\mathcal{U}_{\bm{\sigma}^{\bm{b}},\bm{\sigma}}\psi_{\bm{\lambda\mu}}(\bm{\sigma})$ are calculated directly on the measurement data. Since the negative phase does not depend on the phase parameters $\mu$, standard CD training can be directly applied here. A detailed derivation of the gradients can be found in Ref.~\cite{TorlaiThesis}.

\subsection{Density operators}
When the purity of the quantum state of interest cannot be assumed, one needs to reconstruct the full density operator, $\varrho(\bm{\sigma},\bm{\sigma}^\prime)=\langle\bm{\sigma}|\hat{\varrho}|\bm{\sigma}^\prime\rangle$. Similar to the case of a pure state, before handling the reconstruction we require a representation of the density matrix in terms of a set of network parameters, $\rho_{\bm{\lambda\mu}}(\bm{\sigma},\bm{\sigma}^\prime)$, i.e.~a {\it neural density operator} (NDO). However, in contrast with an RBM wavefunction, the construction of a NDO has more stringent requirements, namely the Hermitian condition $\hat{\rho}_{\bm{\lambda\mu}}=\hat{\rho}^\dagger_{\bm{\lambda\mu}}$ and the positive semi-definite condition $\hat{\rho}_{\bm{\lambda\mu}}\ge0$. One way to enforce the latter directly into the neural network representation consists of adding a set of auxiliary degrees of freedom that purifies the mixed state of the physical system~\cite{Torlai_latent}.

\subsubsection{Latent space purification}
For any mixed quantum state, it is always possible to introduce a set of variables $\bm{\alpha}$ in such a way that the quantum state of the composite system is pure~\cite{Benenti}. In the context of neural networks, we can introduce a RBM wavefunction for the enlarged system
\begin{equation}
|\psi_{\bm{\lambda\mu}}\rangle=\sum_{\bm{\sigma a}}\psi_{\bm{\lambda\mu}}(\bm{\sigma},\bm{\alpha})|\bm{\sigma}\rangle\otimes|\bm{\alpha}\rangle
\end{equation}
and obtain a NDO by tracing out the auxiliary variables:
\begin{equation}
\rho_{\bm{\lambda\mu}}(\bm{\sigma},\bm{\sigma}^\prime)=\sum_{\bm{\alpha}}\psi^*_{\bm{\lambda\mu}}(\bm{\sigma},\bm{\alpha})\psi_{\bm{\lambda\mu}}(\bm{\sigma}^\prime,\bm{\alpha}).
\end{equation}
By embedding the auxiliary units in the latent space of the neural network, it is possible to perform this trace analytically~\cite{Torlai_latent}
\begin{equation}
\rho_{\bm{\lambda\mu}}(\bm{\sigma},\bm{\sigma}^\prime)=\frac{1}{Z_{\bm{\lambda}}}e^{-\Gamma_{\bm{\lambda}}^{[+]}(\bm{\sigma},\bm{\sigma}^\prime)-\Gamma_{\bm{\mu}}^{[-]}(\bm{\sigma},\bm{\sigma}^\prime)-\Pi_{\bm{\lambda\mu}}(\bm{\sigma},\bm{\sigma}^\prime)}\:.
\end{equation}
Here we defined
\begin{equation}
\Gamma_{\bm{\lambda}/\bm{\mu}}^{[\pm]}(\bm{\sigma},\bm{\sigma}^\prime)=\frac{1}{2}\Big[\mathcal{E}_{\bm{\lambda}/\bm{\mu}}(\bm{\sigma})\pm\mathcal{E}_{\bm{\lambda}/\bm{\mu}}(\bm{\sigma}^\prime)\Big]
\end{equation}
and 
\begin{equation}
\Pi_{\bm{\lambda\mu}}(\bm{\sigma},\bm{\sigma}^\prime)=-\sum_k\log\bigg[1+e^{\frac{1}{2}\bm{V}_{\bm{\lambda}}(\bm{\sigma}+\bm{\sigma}^\prime)+\frac{i}{2}\bm{V}_{\bm{\mu}}(\bm{\sigma}-\bm{\sigma}^\prime)}\bigg]\:,
\end{equation}
capturing, respectively, the correlations within the system, and the correlations between the system and the environment. The new parameters $\bm{V}_{\bm{\lambda}/\bm{\mu}}$ encode the degree of mixing of the state of the physical system -- they are identically zero for a pure state.

The cost function for the quantum reconstruction of a NDO is given by
\begin{equation}
\mathcal{C}_{\bm{\lambda\mu}}=-\frac{1}{\|\mathcal{D}\|}\sum_{\bm{\sigma^b}\in\mathcal{D}}\log\rho_{\bm{\lambda\mu}}(\bm{\sigma}^{\bm{b}},\bm{\sigma}^{\bm{b}})-\mathbb{H}_{\mathcal{D}},
\label{Eq::cost_ndo}
\end{equation}
and its gradients can be easily calculated analytically~\cite{TorlaiThesis,Torlai_latent}. Similarly to the case of a pure state, all the gradients can be evaluated directly on the training data (provided the appropriate unitary rotations are applied to the state).  The exception is the term involving the partition function (the negative phase), which is approximated by the CD algorithm using a finite step of block Gibbs sampling (equivalent to sampling the distribution $\rho_{\bm{\lambda\mu}}(\bm{\sigma},\bm{\sigma})$). Given that the purification through the latent space of a RBM architecture generates a physical density operator ($\hat{\rho}_{\bm{\lambda\mu}}\ge0$), this type of ansatz is also suitable for the simulation of quantum dynamics of open systems, which was recently explored in various numerical experiments~\cite{Hartmann,Ciuti,Savona19}.

When evaluating the gradients of the cost function in Eq.~\ref{Eq::cost_ndo}, the NDO needs to be transformed back into the reference basis by appropriate unitary transformations related to the measurement basis, $\hat{\rho}^{\bm{b}}_{\bm{\lambda\mu}}=\hat{\mathcal{U}}_{\bm{b}}\hat{\rho}_{\bm{\lambda\mu}}\hat{\mathcal{U}}^\dagger_{\bm{b}}$. This rotation has to be carried out explicitly and it is then only feasible as long as $\hat{\mathcal{U}}$ acts non-trivially on a sufficiently small number of degrees of freedom. This limitation can be circumvented by avoiding the parametrization of the quantum state directly, and using instead a generative model to represent the probability distribution underlying the measurement outcomes of a set of information complete set of positive-operator valued measures (POVM)~\cite{carrasquilla_povm}. 

\subsection{Reconstruction of experimental wavefunctions}
We have shown that RBMs trained with unsupervised learning offer a versatile approach to quantum state reconstruction of many-body systems. In this Section, we turn to the case of RBM 
reconstruction of experimental data from NISQ hardware.

\begin{figure*}[t]
\includegraphics[width=1.8\columnwidth]{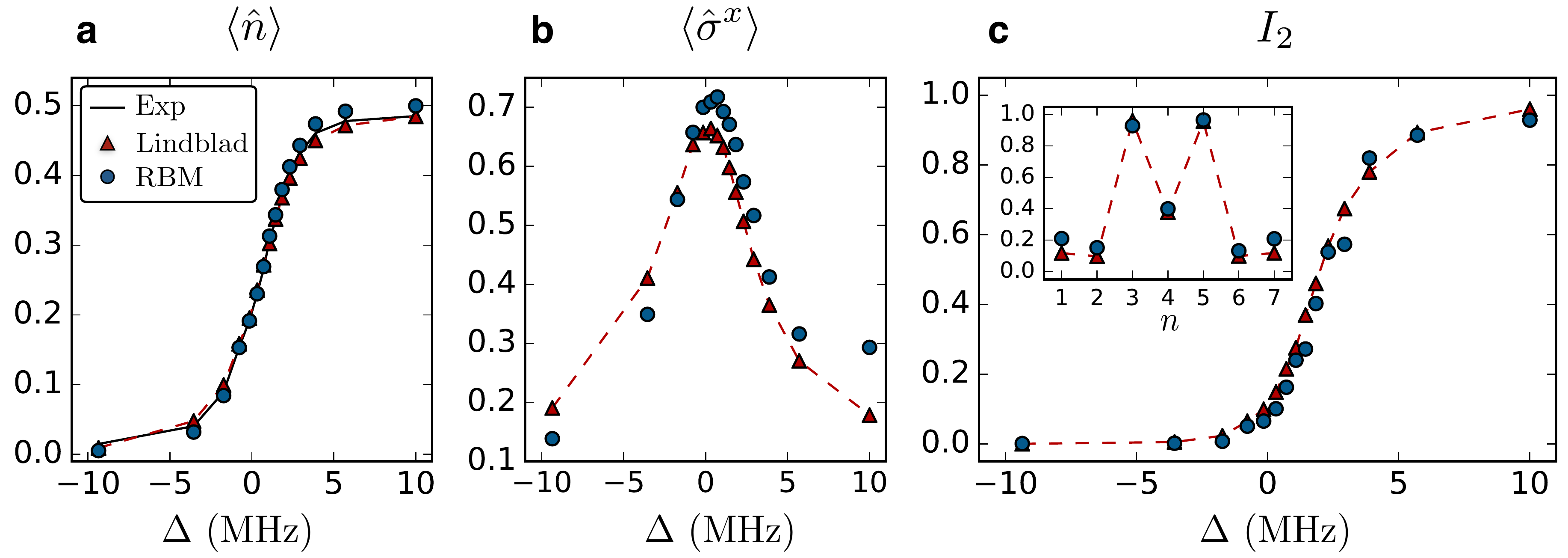}
\caption{Quantum reconstruction of the experimental wavefunction of a Rydberg-atom quantum simulator. ({\it a}) Spatial average of the Rydberg population. ({\it b}) Spatial average of the transverse field. ({\it c}) Renyi mutual information on the bond $(3,4)$ as a function of detuning (main) and in the $\mathbb{Z}_2$ phase as a function of the bond (inset). Figure reproduced from~\cite{torlai19}.}
\label{Fig::rydberg}
\end{figure*}

\subsubsection{Noise mitigation}
One of the major obstacles in reconstructing quantum states from real experiments is the presence of measurement errors. In practice, when performing measurements on a system prepared in the quantum state $\hat{\varrho}$, one obtains measurement outcomes $\bm{\tau}$ which do not correspond to projective measurements, but are instead described by a POVM $\hat{\Pi}^{(\bm{\tau})}=\sum_{\bm{\sigma}}p(\bm{\tau}\:|\:\bm{\sigma})|\bm{\sigma}\rangle\langle\bm{\sigma}|$, where the distribution $p(\bm{\tau}\:|\:\bm{\sigma})$ is the probability of recording the outcome $|\bm{\tau}\rangle$ given the actual measurement $|\bm{\sigma}\rangle$. The probability distribution underlying a set of measurement data is then given by $P(\bm{\tau})=\text{Tr}[\Pi^{(\bm{\tau})}\hat{\varrho}]$. Assuming the rate of measurement errors $p(\bm{\tau}\:|\:\bm{\sigma})$ are known from the experiment, it is possible to incorporate the noisy measurements in the RBM architecture in such a way that the neural network learns the de-noised distribution, corresponding to the  projective measurements prior to the noisy measurement process~\cite{torlai19}.

We show this so-called {\it noise regularization} for the case of a pure and positive RBM wavefunction $\psi_{\bm{\lambda}}(\bm{\sigma})=\sqrt{p_{\bm{\lambda}}(\bm{\sigma})}$, where $p_{\bm{\lambda}}(\bm{\sigma})$ now identifies the probability distribution over the de-noised visible layer $|\bm{\sigma}\rangle$ (before imperfect measurements). Rather than feeding the data directly to this layer, the RBM can be enlarged by adding a third (noise) layer encoding the variable $\bm{\tau}$, with underlying probability distribution $\widetilde{p}_{\bm{\lambda}}(\bm{\tau})=\sum_{\bm{\sigma}}p(\bm{\tau}\:|\:\bm{\sigma})p_{\bm{\lambda}}(\bm{\sigma})$. The weights connecting the noisy and the de-noised layer can be crafted in such a way that the resulting conditional distribution $p(\bm{\tau}\:|\:\bm{\sigma})$ matches the experimental values. The quantum reconstruction is performed by minimizing the negative log-likelihood with respect to the noisy data
\begin{equation}
\mathcal{C}_{\bm{\lambda}}=-\frac{1}{\|\mathcal{D}\|}\sum_{\bm{\tau}\in\mathcal{D}}\log \widetilde{p}_{\bm{\lambda}}(\bm{\tau}),
\end{equation}
with the following update rule,
\begin{equation}
\Delta\bm{\lambda}\propto
\frac{1}{\|\mathcal{D}\|}\sum_{\bm{\tau}\in\mathcal{D}}\big\langle\nabla_{\bm{\lambda}}\mathcal{E}_{\bm{\lambda}}(\bm{\sigma})\big\rangle_{p_{\bm{\lambda}}(\bm{\sigma}|\bm{\tau})}-
\big\langle\nabla_{\bm{\lambda}}\mathcal{E}_{\bm{\lambda}}(\bm{\sigma})\big\rangle_{p_{\bm{\lambda}}(\bm{\sigma})}\:.
\end{equation}
The training is realized similarly to the case of ideal measurements, with the only difference being that the learning in the positive phase, instead of being calculated directly from the data, is driven by the Bayesian posterior distribution
\begin{equation}
p_{\bm{\lambda}}(\bm{\sigma}\:|\:\bm{\tau})=\frac{p(\bm{\tau}\:|\:\bm{\sigma})p_{\bm{\lambda}}(\bm{\sigma})}{\widetilde{p}_{\bm{\lambda}}(\bm{\tau})}\:.
\end{equation}
Following this training procedure, the trained RBM wavefunction $\psi_{\bm{\lambda}}(\bm{\sigma})=\sqrt{p_{\bm{\lambda}}(\bm{\sigma})}$ approximates the quantum state prior to the application of the measurement errors. A similar strategy based on autoencoders has also been put forward, and applied for the neural-network quantum reconstruction of experimental photonic states~\cite{biamonte_qst}.

\subsubsection{Application to a Rydberg-atom quantum simulator}
Finally, we summarize a recent experiment where RBM quantum reconstruction was applied to real data from a NISQ simulator. Specifically, the  experimental system consists of an array of cold Rydberg atoms~\cite{Bernien2017,Endres2016}, one of the highest-quality platforms for programmable simulation of Ising-like quantum spins~\cite{Schauss1455,Labuhn,Sanchez}. In the experiment, $^{87}$Rb atoms are individually trapped by optical tweezers in a defect-free array. The atomic ground state $|g\rangle$ is coupled to an highly excited Rydberg state $|r\rangle$ by a uniform laser drive, and the atoms interact with a Van der Waals potential, resulting into the Hamiltonian
\begin{equation}
\hat H(\Omega, \Delta) = -\Delta \sum_i \hat n_i - \frac{\Omega}{2} \sum_i \hat \sigma_i^x + \sum_{i<j} \frac{V_{nn}}{|i-j|^6} \hat n_i \hat n_j.
\label{Eq::ryd_ham}
\end{equation}
Here, $\Omega$ is the Rabi frequency, $V_{nn}$ is the nearest-neighbour interaction, $\hat{n}_i$ is the occupation number operator for the Rydberg state and $\Delta$ is the laser detuning.
For small ``transverse field'' $\Omega$ and large negative detuning $\Delta$, the ground state of $\hat{H}$ is approximately a product state with all atoms in the ground state $|g\rangle$. Conversely, when the detuning is large and positive, depending on the spacing between the atoms, the simulator can be driven into different broken-symmetry states~\cite{Bernien2017}.

After initializing the simulator in the fiducial state where all atoms are prepared in $|g\rangle$, the ground state of Hamiltonian~(\ref{Eq::ryd_ham}) in the broken-symmetry phase is obtained by an adiabatic ramp in the laser detuning $\Delta$. Assuming perfect adiabatic evolution, the instantaneous state of the system can always be gauged to be positive. Therefore, under this assumption, measurements in the occupation number basis $|\bm{\sigma}\rangle$ (i.e. eigenstates of $\hat{n}_j$) are sufficient to reconstruct the quantum state. The real experimental state is generally expected to be described by a mixed density operator, due to the unavoidable decoherence. Nevertheless, if the purity of the state remains sufficiently high, a pure-state RBM approximation is capable of correctly capturing properties defined over local subsystems~\cite{torlai19}. As described in the previous Section, measurement errors are mitigated using a noise layer regularization.

The experiment is performed with $N=8$ atoms for the transition into the $\mathbb{Z}_2$ ordered phase. The adiabatic sweep is halted at subsequent times (corresponding to an increase laser detuning $\Delta$), where a collection of about 3,000 measurements is taken. Each measurement consists of a bit-string $\bm{\tau}$ recording whether each single atom was measured in the ground or in the Rydberg state. Each dataset is then independently input to an RBM, and used to discover an optimal pure state approximation by incorporating the measurement errors with the noise layer. Once trained, directly sampling the RBM gives access to observable diagonal in the measurement basis. As an example, we show in Fig.~\ref{Fig::rydberg}a the average Rydberg population at different detunings during the sweep. We can compare the values generated by the RBM results with the experimental data (providing direct verification) and with full simulations of the Limbladian master equation. The RBM can then be used to sample expectation values of observables not accessible in the experimental apparatus. In particular, we show the average transverse field (Fig.~\ref{Fig::rydberg}b) and the Renyi mutual information $I_2(s) = S_2(\hat{\rho}_s^A)+S_2(\hat{\rho}_s^B)-S_2(\hat{\rho})$ (Fig.~\ref{Fig::rydberg}c), showing an overall good agreement with the results from simulations of open system dynamics.

\section{Conclusions and outlook}

Modern machine learning has provided us with generative modeling techniques that are perfectly suited for the emerging landscape of NISQ hardware.  
Stochastic neural networks, such as RBMs and their cousins~\cite{carrasquilla_povm,rocchetto,biamonte_qst}, are heuristically known to provide good quality state reconstructions for intermediate-scale and noisy data.  In that sense, their adoption to quantum state reconstruction on devices of tens, hundreds, or even thousands of qubits should come as no surprise.

As discussed in this review, the systematic development of RBM theory for use in quantum state reconstruction is becoming well understood from a formal standpoint. Parallel to theoretical and algorithmic advancements, a crucial role is also played by the development of related open source software~\cite{qucumber,netket}, easily accessible to experimentalists.
However, many fundamental questions still remain to be answered if such machine learning techniques are to become fully integrated with NISQ hardware.

First, as evident in this review, the most well-studied cases involve wavefunctions that are real and positive -- mathematically equivalent to probability distributions.  There,
standard generative models such as RBMs can be employed with little alteration from their original industry-motivated design.  Under the assumption of purity, recent 
work has demonstrated the efficiency of modern algorithms for unsupervised learning in approximate state reconstruction.  In this case, RBMs in particular have shown their utility in producing accurate and scalable estimators for physical observables not directly available from the original data set; i.e.~they generalize well.  Of particular interest is the basis-independent Renyi entanglement entropy, which can be measured directly from a trained RBM using a scalable algorithm involving replication of the model wavefunction.  This is perhaps the most striking example of a measurement that is resource-intensive experimentally~\cite{Islam15}, but relatively simple to implement in the trained generative model.

Real and positive wavefunctions occupy a special place in the landscape of physically-interesting states; for example, they are the ground states of stoquastic Hamiltonians.
However, a large proportion of quantum states under study theoretically and experimentally cannot be assumed to have this significant simplification.  As we have discussed in detail, in the case of complex wavefunctions, state reconstruction is possible with RBMs (and other generative models).  What is required first is a convention to parameterize the phase, e.g.~in additional hidden layers, or as complex weights~\cite{Carleo}.  Then, measurements in more than one basis are needed to train the parameters encoding the phase of the wavefunction.  Given this strategy,
 experimental NISQ wavefunctions, such as cold-atom implementations of the fermionic Hubbard model \cite{Hubbard} or other interesting many-body Hamiltonians, may conceivably be reconstructed in the near future.

Herein lies one frontier for state reconstruction with machine learning.  In the quest to construct a NISQ-compatible generative modeling method, foremost is the question of scaling of the number of measurement bases required for informational completeness.  Very little is known theoretically about this scaling for wavefunctions of interest to NISQ simulators; in the pure case, the number of basis required to learn a $N$-qubit state could range from 1 (see above), to a number that grows exponentially in $N$ (see e.g.~Ref.~\cite{RayL}). This wide range of possibilities leaves open many  questions about the learnability of quantum states.
For example, for what other typical physical wavefunctions is
the number of bases tractable in the context of generative models (RBMs or otherwise)?  Also, how does the target wavefunction structure affect the number of measurements required in each basis?  Finally, what is the relationship between these numbers and the scaling of the RBM parameters required for a desired representational accuracy?
An entire field related to the study of how efficient learning relates to the sign or entanglement structure of a quantum state still lies ahead.

Moving away from pure states, the ability to represent density matrices suggests the possibility that machine-learning reconstruction can be expressed as an approximate re-formulation of more traditional quantum state tomography.  The same scaling questions apply as above for the context of generic complex wavefunctions (albeit with the possibility of significant further roadblocks to scaling).  A reformulation of the problem in the language of informationally-complete POVMs, briefly mentioned here~\cite{carrasquilla_povm}, offers the tantalizing possibility of scaling improvements, at the cost of (experimentally) more complicated measurements.  Finally, success with the mixed-state density matrix formulation suggests
that a re-imagination of process tomography as an unsupervised learning problem could also be in store.
With further development along these lines, generative modeling is poised to breach beyond the realm of NISQ simulators, 
to become a tool for gate-based architectures in the near future.

Looking forward, it is clear that today's hardware is a necessary stepping stone to the more powerful quantum technologies of the future.  As these devices continue to grow, they will develop in lock-step with powerful classical algorithms, to aid in all stages of state preparation, measurement, verification, error correction, and more.  With the dawn of artificial intelligence as the most powerful classical computing paradigm of a generation, it stands to reason that machine learning of quantum many-body states will play a critical role in the NISQ era and beyond.

\subsection*{Acknowledgements }
The Flatiron Institute is supported by the Simons Foundation. R.G.M. is supported by NSERC of Canada, a Canada Research Chair, and the Perimeter Institute for Theoretical Physics. Research at Perimeter Institute is supported through Industry Canada and by the Province of Ontario through the Ministry of Research \& Innovation.

\bibliographystyle{apsrev4-1}
\bibliography{bibliography.bib}

\end{document}